\documentstyle[12pt]{article}

\input{epsf.sty}

\renewcommand{\baselinestretch}{1.4}

\newenvironment{myabstract}[1]{
\renewcommand{\baselinestretch}{1.4}
\vspace*{10mm} \large
\begin{center} {\bf Abstract} \end{center}
\begin{center} \parbox{14cm}{\small #1}}{\end{center}}

\input{psfig.sty}

\begin{document}

\pagestyle{myheadings}


\renewcommand{\baselinestretch}{1.7}

\title{The warp drive: hyper-fast travel \\
\vspace{-5mm} within general relativity.}

\author{Miguel Alcubierre\thanks{Present address: Max Planck Institut
f\"{u}r Gravitationsphysik, Albert Einstein Institut, Schlaatzweg~1,
D-14473~Potsdam, Germany.} \\ Department of Physics and Astronomy,
University of Wales, \vspace{-3mm} \\ College of Cardiff, P.O. Box 913,
Cardiff CF1 3YB, UK. \vspace{4mm} }

\date{}

\maketitle


\begin{myabstract}{

It is shown how, within the framework of general relativity and without
the introduction of wormholes, it is possible to modify a spacetime in
a way that allows a spaceship to travel with an arbitrarily large
speed.  By a purely local expansion of spacetime behind the spaceship
and an opposite contraction in front of it, motion faster than the
speed of light as seen by observers outside the disturbed region is
possible.  The resulting distortion is reminiscent of the ``warp
drive'' of science fiction.  However, just as it happens with
wormholes, exotic matter will be needed in order to generate a
distortion of spacetime like the one discussed here.

\vspace{10mm}

Published in: \hspace{5mm} Class. Quantum Grav. {\bf 11}-5, L73-L77 (1994).

}\end{myabstract}

\pagebreak


\normalsize

When we study special relativity we learn that nothing can travel
faster than the speed of light.  This fact is still true in general
relativity, though in this case one must be somewhat more precise:  in
general relativity, nothing can travel {\em locally\/} faster than the
speed of light.

Since our everyday experience is based on an Euclidean space, it is
natural to believe that if nothing can travel locally faster than light
then given two places that are separated by a spatial proper distance
\,$D$, it is impossible to make a round trip between them in a time
less than \,$2D/c$\, (where \,$c$\, is the speed of light), as measured
by an observer that remains always at the place of departure.  Of
course, from our knowledge of special relativity we know that the time
measured by the person making the round trip can be made arbitrarily
small if his (or her) speed approaches that of light.  However, the
fact that within the framework of general relativity and without the
need to introduce non-trivial topologies (wormholes), one can actually
make such a round trip in an arbitrarily short time as measured by an
observer that remained at rest will probably come as a surprise to many
people.

Here I wish to discuss a simple example that shows how this can be
done.  The basic idea can be more easily understood if we think for a
moment in the inflationary phase of the early Universe, and consider
the relative speed of separation of two comoving observers.  It is easy
to convince oneself that, if we define this relative speed as the rate
of change of proper spatial distance over proper time, we will obtain a
value that is much larger than the speed of light.   This doesn't mean
that our observers will be travelling faster than light:  they always
move inside their local light-cones.  The enormous speed of separation
comes from the expansion of spacetime itself.\,\footnote{\,This
superluminal speed is very often a source of confusion.  It is also a
very good example of how an intuition based on special relativity can
be deceiving when one deals with dynamical spacetimes.}

The previous example shows how one can use an expansion of spacetime to
move away from some object at an arbitrarily large speed.  In the same
way, one can use a contraction of spacetime to approach an object at
any speed.  This is the basis of the model for hyper-fast space travel
that I wish to present here:  create a local distortion of spacetime
that will produce an expansion behind the spaceship, and an opposite
contraction ahead of it.  In this way,  the spaceship will be pushed
away from the Earth and pulled towards a distant star by spacetime
itself.   One can then invert the process to come back to Earth, taking
an arbitrarily small time to complete the round trip.

\vspace{3mm}

I will now introduce a simple metric that has precisely the
characteristics mentioned above.  I will do this using the language of
the 3+1 formalism of general relativity~\cite{MTW,York}, because it
will permit a clear interpretation of the results.  In this formalism,
spacetime is described by a foliation of spacelike hypersurfaces of
constant coordinate time $\,t\,$.  The geometry of spacetime is then
given in terms of the following quantities: the 3-metric
$\,\gamma_{ij}\,$ of the hypersurfaces, the lapse function $\,\alpha\,$
that gives the interval of proper time between nearby hypersurfaces as
measured by the ``Eulerian'' observers (those whose four-velocity is
normal to the hypersurfaces), and the shift vector $\,\beta^i\,$ that
relates the spatial coordinate systems on different hypersurfaces.
Using these quantities, the metric of spacetime can be written
as:\,\footnote{\,In the following greek indices will take the values
(0,1,2,3) and latin indices the values (1,2,3).} \begin{eqnarray} d s^2
&=& - \, d \tau^2 \;\;=\;\; g_{\alpha \beta} \; d x^{\alpha} \, d
x^{\beta} \nonumber \\ &=& - \left( \alpha^2 - \beta_i \, \beta^i
\right) \, d t^2 \,+\, 2 \, \beta_i \, d x^i \, d t \,+\, \gamma_{ij}
\; d x^i \, d x^j \;\; .  \label{pseudodistance}  \end{eqnarray}

Notice that as long as the metric $\,\gamma_{ij}\,$ is positive
definite for all values of \,$t$\, (as it should in order for it to be
a spatial metric), the spacetime is guaranteed to be globally
hyperbolic.  Any spacetime that can be described in the language of the
3+1 formalism will therefore have no closed causal curves.

Let us now assume that our spaceship moves along the $\,x\,$ axis of a
cartesian coordinate system.  We want to find a metric that will
``push'' the spaceship along a trajectory described by an arbitrary
function of time $\,x_s (t)\,$.  A metric that has this property is
given by (\mbox{\,$G \,=\, c \,=\, 1$\,})\,:  \begin{eqnarray} \alpha
&=& 1 \;\; , \label{lapse} \\ \beta^x &=& - \, v_s \left( t \right) \,
f \left( r_s \left( t \right) \right) \;\; , \\ \beta^y &=& \beta^z
\;=\; 0 \;\; , \label{shift} \\ \gamma_{ij} &=& \delta_{ij}
\label{spatial-metric} \;\; , \end{eqnarray}

\noindent where:  \[ v_s \left( t \right) \,=\, \frac{d x_s \left( t
\right) }{d t} \;\; , \hspace{17mm} r_s \left( t \right) \,=\, \left[
\left( \rule{0mm}{4mm} x - x_s \left( t \right) \right)^2 \,+\, y^2
\,+\, z^2 \, \right]^{1/2} \; ,  \]

\noindent and where \,$f$\, is the function:  \begin{equation} f \left(
r_s \right) \,=\, \frac{\tanh \left( \rule{0mm}{4mm} \sigma \, \left(
r_s \,+\, R \right) \right) \,-\, \tanh \left( \rule{0mm}{4mm} \sigma
\, \left( r_s \,-\, R \right) \right) }{ \rule{0mm}{5mm} 2 \, \tanh
\left( \sigma \, R \right)} \;\; , \end{equation}

\noindent with \,$R\,>\,0$\, and \,$\sigma\,>\,0$\, arbitrary
parameters.  Notice that for large \,$\sigma$\, the function \,$f(r)$\,
approaches very rapidly a ``top hat'' function:  \begin{equation}
\lim_{\sigma \rightarrow \infty} \, f \left( r_s \right) \;\;=\;\; \left\{ \begin{array}{l} 1 \hspace{5mm}  {\rm for}
\hspace{4mm} r_s \,\in\, \left[ -R,\,R \right] \;\; , \\ 0 \hspace{5mm}
{\rm otherwise} \;\; .  \end{array} \right.  \end{equation}

\noindent With the above definitions, the metric~(\ref{pseudodistance})
can be rewritten as:  \begin{equation} d s^2 \,=\, - d t^2 \,+\, \left(
\rule{0mm}{4mm} d x - v_s \, f \left( r_s \right) \, d t \right)^2
\,+\, d y^2 \,+\, d z^2 \;\; .  \label{final_metric} \end{equation}

It is easy to understand the geometry of our spacetime from the
previous expressions.  First, from equation~(\ref{spatial-metric}) we
see that the 3-geometry of the hypersurfaces is always flat.  Moreover,
the fact that the lapse is given by \mbox{\,$\alpha \,=\, 1$\,} implies
that the timelike curves normal to these hypersurfaces are geodesics,
i.e., the Eulerian observers are in free fall.  Spacetime, however, is
not flat due to the presence of a non-uniform shift.  Nevertheless,
since the shift vector vanishes for \mbox{\,$r_s \,\gg\, R$\,}, we see
that at any time $\,t\,$ spacetime will be essentially flat everywhere
except within a region with a radius of order $\,R\,$, centered at the
point $\,(x_s(t),\,0,\,0)\,$.

Since the 3-geometry of the hypersurfaces is flat, the information
about the curvature of spacetime will be contained in the extrinsic
curvature tensor $\,K_{ij}\,$.  This tensor describes how the
three-dimensional hypersurfaces are embedded in four-dimensional
spacetime, and is defined as:  \begin{equation} K_{ij} \,=\, \frac{1}{2
\, \alpha} \; \left( D_i \, \beta_j \,+\, D_j \, \beta_i \,-\,
\frac{\partial g_{ij}}{\partial t} \right) \label{extrinsic-def} \;\; ,
\end{equation}

\noindent where $\,D_i\,$ denotes covariant differentiation with
respect to the 3-metric \mbox{$\gamma_{ij}$}.
From the form of \,$\alpha$\,
and \,$\gamma_{ij}$\,,  is not difficult to see that this expression
reduces to:  \begin{equation} K_{ij} \,=\, \frac{1}{2} \, \left(
\rule{0mm}{5mm} \partial_i \, \beta_j \,+\, \partial_j \, \beta_i
\right) \label{extrinsic-simp} \;\; , \end{equation}

The expansion \,$\theta$\, of the volume elements associated with the
Eulerian observers is given in terms of $\,K_{ij}\,$ as:
\begin{equation} \theta \,=\, - \, \alpha \,\,  {\rm Tr} \, K \;\; .
\end{equation}

\noindent From this expression it is not difficult to show that:
\begin{equation} \theta \,=\,  v_s \;\; \frac{x_s}{r_s} \;\; \frac{d
f}{d r_s} \;\; . \label{expansion} \end{equation}

\begin{figure}[h] \def\epsfsize#1#2{0.8#1}
\vspace*{20mm}
\centerline{\epsfbox{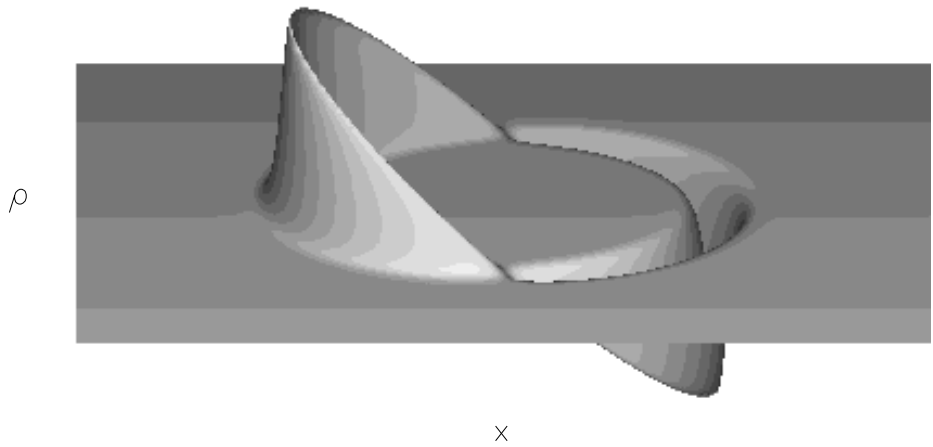}} \vspace{20mm}
\caption{Expansion of the normal volume elements.}
\label{fig:expansion} \end{figure}

Figure\,\ref{fig:expansion} shows a graph of \,$\theta$\, as a function
of \,$x$\, and \mbox{\,$\rho \,=\, (y^2 \,+\, z^2)^{1/2} $}, in the
particular case when \mbox{\,$\sigma \,=\, 8$\,} and \mbox{\,$R \,=\,
v_s \,=\, 1$\,}.  The center of the perturbation \linebreak corresponds
to the spaceship's position \,$x_s \left( t \right)$.  We clearly see
how the volume elements are expanding behind the spaceship, and
contracting in front of it.

To prove that the trajectory of the spaceship is indeed a timelike
curve, regardless of the value of \,$v_s(t)$\,,  we substitute
\mbox{\,$x \,=\, x_s ( t )$\,} in the metric~(\ref{final_metric}).  It
is then easy to see that for the spaceship's trajectory we will have:
\begin{equation} d \tau \,=\, d t \;\; .  \end{equation}

\noindent This implies not only that the spaceship moves on a timelike
curve, but also that its proper time is equal to coordinate time.
Since coordinate time is also equal to the proper time of distant
observers in the flat region, we conclude that the spaceship suffers no
time dilation as it moves. It is also straightforward to prove that the
spaceship moves on a geodesic.  This means that even though the
coordinate acceleration can be an arbitrary function of time, the
proper acceleration along the spaceship's path will always be zero.
Moreover, it is not difficult to convince oneself that when the
parameter \,$\sigma$\, is large, the tidal forces in the immediate
vicinity of the spaceship are very small (provided that \,$R$\, is
larger than the size of the spaceship).  Of course, in the region where
\mbox{\,$r_s \,\simeq\, R$\,} the tidal forces can be very large
indeed.

\vspace{3mm}

To see how one can use this metric to make a round trip to a distant
star in an arbitrary small time, let us consider the following
situation:  Two stars \,$A$\, and \,$B$\, are separated by a distance
\,$D$\, in flat spacetime.  At time \,$t_0$\,, a spaceship starts to
move away from \,$A$\, at a speed \mbox{\,$v < 1 $\,} using its rocket
engines.  The spaceship then stops at a distance \,$d$\, away from
\,$A$\,.  I will assume that \,$d$\, is such that:  \begin{equation} R
\,\ll\, d \,\ll\, D \;\; . \label{condition} \end{equation}

It is at this point that a disturbance of spacetime of the type
described, centered at the spaceship's position, first appears.  This
disturbance is such that the spaceship is pushed away from \,$A$\, with
a coordinate acceleration that changes rapidly from \,$0$\, to a
constant value \,$a$\,.  Since the spaceship is initially at rest
(\mbox{\,$v_s \,=\, 0$\,}), the disturbance will develop smoothly from
flat spacetime (see equation~(\ref{final_metric})).

When the spaceship is halfway between \,$A$\, and \,$B$\,, the
disturbance is modified in such a way that the coordinate acceleration
changes rapidly from \,$a$\, to \,$-a$\,.  If the coordinate
acceleration in the second part of the trip is arranged in such a way
as to be the opposite to the one we had in the first part, then the
spaceship will eventually find itself at rest at a distance \,$d$\,
away from \,$B$\,, at which time the disturbance of spacetime will
disappear (since again \mbox{\,$v_s \,=\, 0$\,}).  The journey is now
completed by moving again through flat spacetime at a speed
\,$v$.\,\footnote{The two constant-velocity legs at the beginning and
end of the journey are not crucial for the argument that I wish to
present here.  I only introduce them in order to guarantee that the two
stars will remain unaffected by the disturbance of spacetime
(\,$R\,\ll\,d$\,), and can therefore be used as unperturbed ``clocks''
with which to compare the proper time on board the spaceship.}

If each of the changes in acceleration are very rapid, the total
coordinate time \,$T$\, elapsed in the one way trip will be essentially
given by:  \begin{equation} T \,=\, 2 \, \left[ \, \frac{d}{v} \,+\,
\sqrt{\frac{D \,-\, 2 d}{a}} \;\, \right] \;\; .
\label{coordinate-time} \end{equation}

\noindent Since both stars remain in flat space, their proper time is
equal to coordinate time.  The proper time measured on the spaceship,
on the other hand, will be:  \begin{equation} \tau \,=\, 2 \, \left[ \,
\frac{d}{\gamma \, v} \,+\, \sqrt{\frac{D \,-\, 2 d}{a}} \;\, \right]
\, , \label{proper-time} \end{equation}

\noindent with \mbox{\,$\gamma \,=\, \left( \rule{0mm}{4mm} 1 - v^2
\right)^{-1/2}$}.  We see then that the time dilation comes only from
the initial and final stages of the trip, when the spaceship moves
through flat spacetime.  Now, if condition~(\ref{condition}) holds, we
will have:  \begin{equation} \tau \,\simeq\, T \,\simeq\, 2 \,\,
\sqrt{\frac{D}{a}} \label{time} \end{equation}

\noindent It is now clear that \,$T$\, can be made as small as we want
by increasing the value of~\,$a$\,.  Since a round trip will only take
twice as long, we find that we can be back in star~\,$A$\, after an
arbitrarily small proper time, both from the point of view of the
spaceship and from the point of view of the star.  The spaceship will
then be able to travel much faster than the speed of light.  However,
as we have seen, it will always remain on a timelike trajectory,  that
is, inside its local light-cone:  light itself is also being pushed by
the distortion of spacetime.  A  propulsion mechanism based on such a
local distortion of spacetime just begs to be given the familiar name
of the ``warp~drive'' of science fiction.

The metric I have just described has one important drawback, however:
it violates all three energy conditions (weak, dominant and
strong~\cite{Hawking}). Both the weak and the dominant energy
conditions require the energy density to be positive for {\em all\/}
observers.  If one calculates the Einstein tensor from the
metric~(\ref{final_metric}), and uses the fact that the four-velocity
of the Eulerian observers is given by:  \begin{equation} n^{\alpha}
\,=\, \frac{1}{\alpha} \, \left( 1 \, , - \beta^i \right) \;\; ,
\hspace{10mm} n_{\alpha} \,=\, - \left( \rule[0mm]{0mm}{4mm} \alpha \,
, \, 0 \right) \;\; , \end{equation}

\noindent then one can show that these observers will see an energy
density given by:  \begin{equation} T^{\alpha \beta} \,  n_{\alpha} \,
n_{\beta} \,=\, \alpha^2 \; T^{\,0\,0} \,=\, \frac{1}{8 \pi} \;\;
G^{\,0\,0} \;=\, - \, \frac{1}{8 \pi} \;\; \frac{v_s^2 \, \rho^2}{4 \,
{r_c}^2} \; \left( \rule{0mm}{4mm} \frac{d f}{d r_s} \right)^2 \;\; .
\end{equation}

\noindent The fact that this expression is everywhere negative implies
that the weak and dominant energy conditions are violated.  In a
similar way one can show that the strong energy condition is also
violated.

We see then that, just as it happens with wormholes, one needs exotic
matter to travel faster than the speed of light.  However, even if one
believes that exotic matter is forbidden classically, it is well known
that quantum field theory permits the existence of regions with
negative energy densities in some special circumstances (as, for
example, in the Casimir effect~\cite{DeWitt}).  The need of exotic
matter therefore doesn't necessarily eliminate the possibility of
using a spacetime distortion like the one described above for
hyper-fast interstellar travel.

As a final comment, I will just mention the fact that even though the
spacetime described by the metric~(\ref{final_metric}) is globally
hyperbolic, and hence contains no closed causal curves, it is probably
not very difficult to construct a spacetime that does contain such
curves using a similar idea to the one presented here.

\vspace{10mm}

The author wishes to thank Bernard F. Schutz and Gareth S. Jones for
many useful comments.


\pagebreak






\large

\end{document}